\documentclass[journal,a4paper]{IEEEtran}

\IEEEoverridecommandlockouts

\usepackage{color}
\usepackage{svg}
\newcommand{\Arrow}[1]{%
\parbox{#1}{\tikz{\draw[->](0,0)--(#1,0);}}
}
\graphicspath{{Figs/}}
\usepackage[nolist]{acronym}
\usepackage{tikzit}

\tikzstyle{square}=[fill=none, draw=black, shape=rectangle, align=center]
\tikzstyle{Text}=[fill=none, draw=none, shape=circle, align=center]

\tikzstyle{arrow}=[->]
\tikzstyle{Thick}=[-, very thick]
\tikzstyle{sts}=[-, fill={rgb,255: red,240; green,177; blue,115}, draw=black, thin]
\tikzstyle{lts}=[-, fill={rgb,255: red,249; green,212; blue,101}, thin]
\tikzstyle{signal}=[-, fill={rgb,255: red,169; green,202; blue,154}, thin]
\tikzstyle{data}=[-, fill={rgb,255: red,127; green,182; blue,234}, thin]
\tikzstyle{double stop}=[{|-|}]
\tikzstyle{stop arrow}=[{|->}]
\tikzstyle{dotted line}=[-, draw={rgb,255: red,128; green,128; blue,128}, dashed]
\tikzstyle{greyed}=[-, fill={rgb,255: red,128; green,128; blue,128}, thin]

\everymath=\expandafter{\the\everymath\displaystyle}

\usepackage{graphicx}
\usepackage{multicol}
\usepackage{multirow}
\usepackage{blindtext}
\usepackage{amsmath}
\usepackage{amssymb}
\usepackage{fancyhdr}
\usepackage{cite}
\usepackage[size=smaller]{caption}
\usepackage{array}                  

\usepackage{booktabs} 
\usepackage{tikz}
\usepackage[utf8]{inputenc}
\usepackage{pgf}
\usepackage{pgfplots}
\usetikzlibrary{plotmarks}
\usepackage{pdflscape}
\pgfplotsset{compat=newest}
\pgfplotsset{plot coordinates/math parser=false}
\newlength\fwidth 

\usepackage{enumitem} 
\usepackage{algorithmic}
\usepackage[ruled,vlined]{algorithm2e}

\usepackage{textcomp}
\usepackage{subcaption}
\usepackage{hyperref}
\usepackage{cleveref}
\crefname{figure}{Fig.}{Figs.}
\Crefname{figure}{Fig.}{Figs.}
\crefname{table}{Table}{Tables}
\Crefname{table}{Table}{Tables}
\crefname{equation}{}{}
\Crefname{equation}{Equation}{Equations}

\usepackage{lmodern}
\usepackage{xcolor}
\usepackage[export]{adjustbox}

\begin{document}

\title{Stealing AI Model Weights Through Covert Communication Channels \vspace{-0.0cm}}



\author{
\IEEEauthorblockN{Valentin Barbaza, Alán Rodrigo Díaz-Rizo, Hassan Aboushady, Spyridon Raptis, \\and Haralampos-G. Stratigopoulos\\}
\IEEEauthorblockA{\small{Sorbonne Universit\'{e}, CNRS, LIP6, Paris, France}}\thanks{This work was funded by the Chips JU project Resilient Trust of the EU’s Horizon Europe research and innovation programme under Grant agreement N$^{\mbox{\scriptsize o}}$ 101112282 and by the EU Network of Excellence dAIEDGE under Grant agreement N$^{\mbox{\scriptsize o}}$ 101120726.}
}

\maketitle

\begin{abstract}
AI models are often regarded as valuable intellectual property due to the high cost of their development, the competitive advantage they provide, and the proprietary techniques involved in their creation. As a result, AI model stealing attacks pose a serious concern for AI model providers. In this work, we present a novel attack targeting wireless devices equipped with AI hardware accelerators. The attack unfolds in two phases. In the first phase, the victim’s device is compromised with a hardware Trojan (HT) designed to covertly leak model weights through a hidden communication channel, without the victim realizing it. In the second phase, the adversary uses a nearby wireless device to intercept the victim’s transmission frames during normal operation and incrementally reconstruct the complete weight matrix. The proposed attack is agnostic to both the AI model architecture and the hardware accelerator used. We validate our approach through a hardware-based demonstration involving four diverse AI models of varying types and sizes. We detail the design of the HT and the covert channel, highlighting their stealthy nature. Additionally, we analyze the impact of bit error rates on the reception and propose an error mitigation technique.
The effectiveness of the attack is evaluated based on the accuracy of the reconstructed models with stolen weights and the time required to extract them. Finally, we explore potential defense mechanisms.

\end{abstract}


\section{Introduction}
\label{sec:intro}

AI models are regarded as valuable assets because their development demands significant investment in data collection, computational resources, and training time. They also offer a competitive edge, as model performance frequently distinguishes companies in the same industry. Furthermore, these models embody proprietary insights, including specialized feature engineering, architectural decisions, and unique training methodologies. As a result, AI model stealing attacks have emerged \cite{OlMaRa23,PoKu24}. They can be broadly classified into two categories: software-based and hardware-based. 

Software-based attacks primarily involve query-based approaches through APIs, without requiring physical access to the AI hardware accelerator and interacting with it as a black box \cite{TZJRR16}. By sending inputs to the model and analyzing the corresponding outputs, they attempt to reconstruct a functionally equivalent surrogate model.

On the other hand, hardware-based attacks require physical access or control over the device. The attacker must have possession of the device  and be able to interact with it in a controlled lab environment using probing or measurement equipment. These attacks can be broadly classified into side-channel attacks (SCA) \cite{WLLLX18,DuCaAy20,YKOSF20,XCCFHCLWXY20,DKAA22, GQZWMAXFN24, BBJP19, DeepSniffer,YMYZJ20,HuZhSu18,DSRB18,GoFeWa20,HDKLRKD-SD18,YaFlTo20, RCYF22,NNQA-G18, JMKM20,WZZLA-F20,ZCZL21,NNQG21}, fault injection attacks (FIA) \cite{BJHBL22, HMDD23}, and scan-based attacks \cite{PoAy21}. SCAs exploit information leakage from the physical hardware during inference, i.e., such as by monitoring power consumption 
\cite{WLLLX18, DuCaAy20,YKOSF20,XCCFHCLWXY20,DKAA22,GQZWMAXFN24}, measuring 
electromagnetic (EM) radiation \cite{BBJP19, DeepSniffer,YMYZJ20}, exploiting variations in execution time \cite{HuZhSu18,DSRB18,GoFeWa20}, analyzing memory usage patterns \cite{HDKLRKD-SD18, YaFlTo20, RCYF22} or targeting platform-specific, e.g., GPU, leaks \cite{NNQA-G18, JMKM20, WZZLA-F20, ZCZL21,NNQG21}.
In FIA, the attacker injects faults to cause abnormal behavior and extract information. In scan-based attacks, the attacker leverages the scan test infrastructure to read internal states and retrieve model data, such as weights.

In this work, we introduce a novel hardware-based attack designed to reverse-engineer AI model parameters from a hardware device via an RF-based covert channel.
We assume that the device includes at least wireless communication capabilities and an AI hardware accelerator with on-chip memory for storing model parameters. The covert channel leaks the model parameters within legitimate wireless transmissions, remaining undetectable by the devices involved and without affecting their communication. The attacker introduces a malicious modification to the hardware, e.g. a hardware Trojan (HT), to enable the covert channel. While the victims upload the AI model onto the device to perform inference, unbeknownst to them, the device is gradually disclosing the model parameters to the attacker during communication with other devices via the covert channel. The attacker could be either the device provider or the foundry that manufactured the device or a third-party attacker who commissions the provider or the foundry to facilitate the attack. The proposed attack is generic and applicable to any type of AI model or AI hardware accelerator architecture.

In contrast to SCAs, the proposed attack does not interfere with the device's operation, meaning it does not require querying, probing, or measuring the device. The attacker does not need physical possession of the device; only needs to be within the communication range. Therefore, while in a SCA the attacker steals a model by controlling the device, in the proposed attack, the device secretly leaks the model concurrently with its operation, without the user being aware of it.

The complete attack unfolds in the following stages: (a) the attacker inserts a HT into the AI hardware accelerator device, establishing a covert communication channel; (b) this covert channel is specifically designed to leak the weight matrix of the AI model stored within the device; (c) during regular operation of the AI hardware accelerator device, the attacker is positioned in wireless range and, without the user's awareness, retrieves the transmitted data and gradually extracts and steals the weights, ultimately recovering the entire AI model.

Various approaches for establishing covert channels have been proposed, targeting different layers of the wireless communication stack. These include the Medium Access Control (MAC) protocol \cite{KKCR13}, the digital baseband physical (PHY) layer \cite{Dutta13, ClScHo15,HiFr10,GrSz13,SAANM19,DiRiAbSt22}, or the analog front-end of the RF transceiver \cite{JiMa10,LJNM17,SHANM20, CBOBO18,SRDJWA-SDMIC19}. 
Such techniques are generally employed to leak cryptographic keys, enabling decryption of future communications. In contrast, this work introduces—for the first time—the use of covert channels to exfiltrate AI models from edge devices. In our implementation, we use a state-of-the-art covert channel proposed in \cite{DiRiAbSt22}, but with a completely different and hardware-efficient realization than the one described in \cite{DiRiAbSt22}.

We make the following contributions:

\begin{itemize}
    \item  We present the first hardware demonstration of AI model leakage from a chip running the model via an RF-based covert channel,  representing a conceptually distinct attack approach from conventional SCA. 
    \item We show how to establish the covert channel using a stealthy HT with minimal overhead. While we rely on the approach proposed in \cite{DiRiAbSt22} that hides the stolen bits of information into the preamble of the transmission frame, we propose an entirely different and more efficient hardware implementation of the HT that operates in the time-domain, as opposed to the frequency-based design in \cite{DiRiAbSt22}.
    \item While HT design is a well-established field \cite{BHBN14,XFJKBT16}, and prior work has explored HTs for inducing denial-of-service in AI hardware accelerators via input triggers \cite{RKKAS25}, this is the first work to leverage a HT specifically for leaking AI models.
    \item The proposed attack is agnostic to both the AI hardware accelerator and the neural network model, as the HT simply sniffs weight values stored in memory without interfering with the accelerator’s internal computations. In fact, we demonstrate the attack across a range of models, including image classification networks (like LeNet-5 and MobileNetV3-Large), an object detection model (YOLOv11n), and a Spiking Neural Network (SNN) trained on IBM’s hand and arm gesture recognition dataset.
    \item In our Wi-Fi covert channel hardware demonstration, we account for practical communication constraints by varying the Signal-to-Noise Ratio (SNR) of the channel, which affects the Bit Error Rate (BER) at reception.
    We show that BER-induced bit flips in the model's weight matrix can degrade the accuracy of the leaked model and analyze, for each model in our benchmarks, the minimum BER required to recover baseline accuracy. Under less favorable BER conditions, we quantify how many repetitions of the leakage are needed to mitigate errors through bit voting and correction, effectively restoring model accuracy. Our results show that, under a stable high-speed Wi-Fi connection, even the largest model in our benchmarks can be reliably leaked within two hours. Additionally, we present trade-off curves illustrating the relationship between BER, leakage repetitions, and resulting model accuracy.
\end{itemize}

The rest of the article is structured as follows. Section \ref{sec:threat_model} introduces the threat model. Section \ref{sec:attack_principle} provides an overview of the attack principle. Section \ref{sec:covert_channel} describes the covert channel technique. Section \ref{sec:implementation} details the hardware implementation. Section \ref{sec:case_studies} discusses the case studies, and Section \ref{sec:results} reports the experimental results. Finally, Section \ref{sec:conclusion} concludes the article.
\section{Threat Model}
\label{sec:threat_model}

The AI model weights are leaked by an edge device that integrates both an AI hardware accelerator and wireless communication capabilities. The adversary may be the design house of the device or the foundry to which fabrication is outsourced, capable of modifying the device to insert the HT. Alternatively, the adversary could be a third-party attacker who commissions the design house or the foundry to facilitate the attack.

The victim possess a HT-infected device with the HT enabling a covert channel for leaking the AI model's parameters. The victim loads the AI model onto the device to accelerate inference, unaware that, during operation, the HT inconspicuously exfiltrates the parameters via a covert channel embedded within legitimate transmissions. The parameters are stored as bits in an on-chip memory, and the covert channel is leaking a number of bits per transmission frame.

We assume that the adversary has grey-box access, that is, the adversary knows the AI model architecture and hyperparameters (i.e., number of layers, layer type, layer connectivity, feature map size, convolution operations, etc.) but not the learned parameters, i.e., weights. The goal of the adversary is to steal these parameters, which represents the asset and valuable intellectual property of the victim. Another incentive for the adversary is to recover the model so as to craft adversarial examples to perform an evasion attack, i.e., subtly manipulate the input to fool the model at inference time, without changing the model itself \cite{SZSBEG14}.

\begin{figure}
    \centering
    \resizebox{.52\columnwidth}{!}{\includesvg[width=.6\columnwidth]{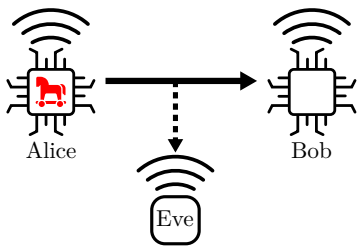}}
    \caption{Attack threat model.}\vspace{-0.0cm}
    \label{fig:attack-model}
\end{figure}

In our implementation we demonstrate the covert channel for Wi-Fi communication. The victim, e.g. Alice, is transmitting data into the surrounding space which are picked by a Wi-Fi access point or another Wi-Fi-enabled device, e.g., Bob, as illustrated in \cref{fig:attack-model}. The adversary, e.g. Eve, is a Wi-Fi-enabled device equipped with an RF receiver capable of receiving and processing the transmitted frames by Alice. Eve needs to be at Wi-Fi range with Alice, that is, the distance between Eve and Alice should be within the typical operating range of Wi-Fi communication. The effective range depends on various factors, i.e., walls, interference, and antenna quality. For Wi-Fi operating in the 2.4 GHz band, this range is roughly 30–50 meters indoors and 100+ meters outdoors. A signal strong enough for communication at typical Wi-Fi power levels has a signal-to-noise ratio (SNR) of around 20 dB. Eve, once at Wi-Fi range with Alice, can begin receiving the transmitted frames, each of which contains a portion of the AI model's parameters, ultimately collecting the full set of parameters.

Although the attack is demonstrated for Wi-Fi, it virtually applies to other communication protocols too, such as Bluetooth and ZigBee.

\section{Attack principle}
\label{sec:attack_principle}

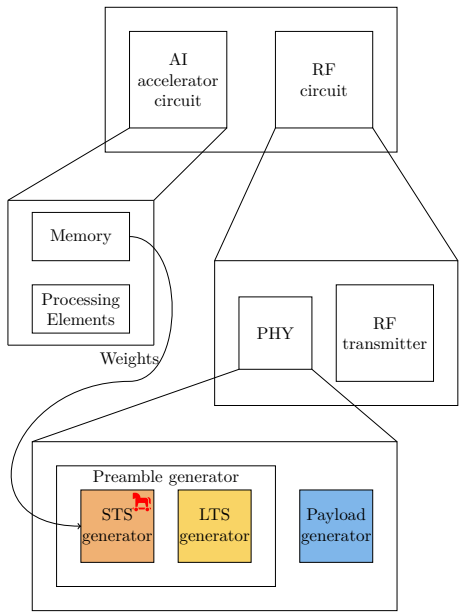
\begin{figure}
    \centering
    \resizebox{.8\linewidth}{!}{\begin{tikzpicture}
  \begin{pgfonlayer}{nodelayer}
    \node [style=none] (6) at (0.5, -0.5) {};
    \node [style=none] (7) at (2.5, -0.5) {};
    \node [style=none] (8) at (0.5, -2.5) {};
    \node [style=none] (9) at (2.5, -2.5) {};
    \node [style=none] (22) at (3.5, -0.5) {};
    \node [style=none] (23) at (5.5, -0.5) {};
    \node [style=none] (24) at (3.5, -2.5) {};
    \node [style=none] (25) at (5.5, -2.5) {};
    \node [style=none] (29) at (0, 0) {};
    \node [style=none] (30) at (6, 0) {};
    \node [style=none] (31) at (0, -3) {};
    \node [style=none] (32) at (6, -3) {};
    \node [style=Text] (33) at (1.5, -1.5) {AI\\accelerator\\circuit};
    \node [style=Text] (34) at (4.5, -1.5) {RF\\circuit};
    \node [style=none] (35) at (-2, -7) {};
    \node [style=none] (36) at (1, -7) {};
    \node [style=none] (37) at (-2, -4) {};
    \node [style=none] (38) at (1, -4) {};
    \node [style=none] (39) at (-1.5, -5.25) {};
    \node [style=none] (40) at (0.5, -5.25) {};
    \node [style=none] (41) at (-1.5, -4.25) {};
    \node [style=none] (42) at (0.5, -4.25) {};
    \node [style=none] (43) at (-1.5, -6.75) {};
    \node [style=none] (44) at (0.5, -6.75) {};
    \node [style=none] (45) at (-1.5, -5.75) {};
    \node [style=none] (46) at (0.5, -5.75) {};
    \node [style=Text] (47) at (-0.5, -4.75) {Memory};
    \node [style=Text] (48) at (-0.5, -6.25) {Processing\\Elements};
    \node [style=none] (49) at (2.25, -8.25) {};
    \node [style=none] (50) at (7.25, -8.25) {};
    \node [style=none] (51) at (2.25, -5.25) {};
    \node [style=none] (52) at (7.25, -5.25) {};
    \node [style=none] (53) at (2.75, -6) {};
    \node [style=none] (54) at (4.25, -6) {};
    \node [style=none] (55) at (2.75, -7.5) {};
    \node [style=none] (56) at (4.25, -7.5) {};
    \node [style=Text] (57) at (3.5, -6.75) {PHY};
    \node [style=none] (58) at (4.75, -5.75) {};
    \node [style=none] (59) at (6.75, -5.75) {};
    \node [style=none] (60) at (4.75, -7.75) {};
    \node [style=none] (61) at (6.75, -7.75) {};
    \node [style=Text] (62) at (5.75, -6.75) {RF\\transmitter};
    \node [style=none] (63) at (-1.5, -12.5) {};
    \node [style=none] (64) at (6, -12.5) {};
    \node [style=none] (65) at (-1.5, -9) {};
    \node [style=none] (66) at (6, -9) {};
    \node [style=none] (67) at (-0.5, -10) {};
    \node [style=none] (68) at (1, -10) {};
    \node [style=none] (69) at (-0.5, -11.5) {};
    \node [style=none] (70) at (1, -11.5) {};
    \node [style=Text] (71) at (0.25, -10.75) {STS\\generator};
    \node [style=none] (72) at (1.5, -10) {};
    \node [style=none] (73) at (3, -10) {};
    \node [style=none] (74) at (1.5, -11.5) {};
    \node [style=none] (75) at (3, -11.5) {};
    \node [style=Text] (76) at (2.25, -10.75) {LTS\\generator};
    \node [style=none] (77) at (4, -10) {};
    \node [style=none] (78) at (5.5, -10) {};
    \node [style=none] (79) at (4, -11.5) {};
    \node [style=none] (80) at (5.5, -11.5) {};
    \node [style=Text] (81) at (4.75, -10.75) {Payload\\generator};
    \node [style=none] (82) at (-0.5, -10.75) {};
    \node [style=none] (83) at (0.5, -4.75) {};
    \node [style=Text, label={above:Weights}] (84) at (0.5, -7.75) {};
    \node [style=none] (85) at (-1, -9.5) {};
    \node [style=none] (86) at (-1, -12) {};
    \node [style=none] (87) at (3.5, -12) {};
    \node [style=none] (88) at (3.5, -9.5) {};
    \node [style=Text] (89) at (1.25, -9.75) {};
    \node [style=Text] (90) at (1.25, -9.75) {Preamble generator};
    
    \node [style=none] (trojan) at (0.75, -10.25) {\resizebox{\linewidth}{!}{\def \globalscale {0.000700}
\begin{tikzpicture}[y=1px, x=1px, yscale=\globalscale,xscale=\globalscale, every node/.append style={scale=\globalscale}, inner sep=0pt, outer sep=0pt]
  \begin{scope}[fill=red]
    \path[fill=red] (458.5, 100.0) -- (437.5,
    100.0)arc(-44.19999999999999:0.1:64.4 and -64.4) -- (458.5,
    55.0)arc(90.0:0.0:22.5 and -22.5) -- (481.0, 77.5)arc(0.0:-90.0:22.5 and
    -22.5) -- cycle;

    \path[fill=red] (391.0, 55.1) circle (45.0px);

    \path[fill=red] (185.8, 55.0) -- (326.2, 55.0)arc(179.9:224.3:64.4 and
    -64.4) -- (167.5, 100.0)arc(-44.19999999999999:0.1:64.4 and -64.4) -- cycle;

    \path[fill=red] (121.0, 55.1) circle (45.0px);

    \path[fill=red] (31.0, 77.5) -- (31.0, 77.5)arc(180.0:90.0:22.5 and -22.5)
    -- (56.2, 55.0)arc(179.9:224.3:64.4 and -64.4) -- (53.5,
    100.0)arc(270.0:180.0:22.5 and -22.5) -- cycle;

    \path[fill=red] (428.8, 280.1) -- (428.8, 215.3) -- (456.6,
    187.5)arc(135.0:0.0:9.0 and -9.0) -- (472.0, 286.7).. controls (472.0, 316.7)
    and (452.1, 343.9) .. (422.9, 350.5)arc(282.9:250.6:62.2 and
    -62.2)arc(298.9:359.9:78.3 and -78.3) -- cycle;

    \path[fill=red] (103.0, 117.2)arc(253.3:270.2:62.2 and
    -62.2)arc(270.0:297.6:64.8 and -64.8) -- (170.9, 195.3)arc(193.5:270.0:22.5
    and -22.5) -- (319.2, 212.5)arc(270.0:346.5:22.5 and -22.5) -- (361.0,
    112.5)arc(242.4:270.0:64.7 and -64.7)arc(269.8:286.7:62.2 and -62.2) --
    (409.0, 280.0)arc(0.0:-90.0:58.5 and -58.5) -- (224.5,
    338.5)arc(0.0:-90.0:121.5 and -121.5) -- (62.5, 460.0)arc(270.0:180.0:22.5 and
    -22.5) -- (40.0, 311.5) -- (76.0, 311.5) -- (103.0, 338.5) -- cycle;

  \end{scope}

\end{tikzpicture}}};
  \end{pgfonlayer}
  \begin{pgfonlayer}{edgelayer}
    \draw (9.center)
    to (8.center)
    to (6.center)
    to (7.center)
    to cycle;
    \draw (23.center)
    to (22.center)
    to (24.center)
    to (25.center)
    to cycle;
    \draw (32.center)
    to [in=0, out=180] (31.center)
    to (29.center)
    to (30.center)
    to cycle;
    \draw (36.center) to (35.center);
    \draw (37.center) to (35.center);
    \draw (36.center) to (38.center);
    \draw (37.center) to (38.center);
    \draw (42.center)
    to (40.center)
    to (39.center)
    to (41.center)
    to cycle;
    \draw (46.center)
    to (44.center)
    to (43.center)
    to (45.center)
    to cycle;
    \draw (8.center) to (37.center);
    \draw (9.center) to (38.center);
    \draw (49.center)
    to (51.center)
    to (52.center)
    to (50.center)
    to cycle;
    \draw (55.center)
    to (53.center)
    to (54.center)
    to (56.center)
    to cycle;
    \draw (24.center) to (51.center);
    \draw (25.center) to (52.center);
    \draw (61.center)
    to (60.center)
    to (58.center)
    to (59.center)
    to cycle;
    \draw (64.center)
    to (63.center)
    to (65.center)
    to (66.center)
    to cycle;
    \draw [style=sts] (67.center)
    to (68.center)
    to (70.center)
    to (69.center)
    to cycle;
    \draw [style=lts] (73.center)
    to (75.center)
    to (74.center)
    to (72.center)
    to cycle;
    \draw [style=data] (78.center)
    to (77.center)
    to (79.center)
    to (80.center)
    to cycle;
    \draw (55.center) to (65.center);
    \draw (56.center) to (66.center);
    \draw [style=arrow] (83.center)
    to [in=0, out=0] (84.center)
    to [in=-180, out=180, looseness=2.00] (82.center);
    \draw [style=none] (87.center)
    to (88.center)
    to (85.center)
    to (86.center)
    to cycle;
  \end{pgfonlayer}
\end{tikzpicture}}
    \caption{Architecture of leaking edge device.}\vspace{-0.0cm}
    \label{fig:soc-overall}
\end{figure}

The attack targets integrated circuits (ICs) that comprise at minimum an AI hardware accelerator and an RF transceiver, as illustrated in \cref{fig:soc-overall}. 

In the first phase of the attack, the adversary inserts the HT into the device, which establishes a covert channel for gradually leaking the model weights. The HT is composed of a trigger and a payload mechanism. In our implementation, the trigger is permanently active, causing the weights to be continuously leaked in a loop—once the full set of weights is transmitted, the leaking cycle restarts. Alternatively, the HT could be designed to be activated only in response to a specific input trigger. 

The weights are stored as digital words in on-chip memory. The payload mechanism comprises two parts. The first establishes a connection from the memory to the RF transmitter, enabling access to the stored weights to sniff them. The second part is located inside the RF transmitter and is responsible for creating the covert channel and leaking the weights through it. In our implementation, the second part is located in the physical (PHY) layer of the RF transceiver, in particular into the preamble generation block, as it will be explained in detail in Section \ref{sec:covert_channel}. In an IC implementation where all components are integrated onto the same substrate, this scheme where secret information, e.g., the weights, in one part of the design is driven to another part of the design, i.e., the PHY layer of the RF transceiver, is entirely feasible if the adversary is the design house or the foundry.

As illustrated in \cref{fig:attack-model}, the compromised device, Alice, transmits frames that covertly leak $B$ bytes of model weights per frame. The adversary, Eve, is at communication range with Alice, intercepts these transmissions and incrementally reconstructs the full weight matrix. Given an AI model with $N_w$ weights represented at $p$-bit precision, the entire model is leaked after Alice transmits 

\begin{equation}
    N_f=\frac{N_w \times p}{8 \times B}
\end{equation}

\noindent frames. A detailed analysis of the resulting throughput in our implementation is provided in \cref{subsec:covert_channel_throughput}. Clearly, the larger the model size and the higher the data precision are, the longer the leakage time will be.

Ideally, the goal of the attack is to steal the exact weights. Eve obtains $N_w \times p$ bits by receiving the transmission from Alice over the air.
As with any wireless communication system, there is a BER resulting from various factors such as noise in Alice's device, channel impairments, interference, and synchronization issues between Alice and Eve. As a result, Eve will reconstruct an approximate weight matrix due to the BER, meaning that some of the weights may have incorrect values. Fault injection experiments, particularly those involving bit flips in memory storing the weights, have demonstrated that AI models are quite resilient to such bit flips \cite{SuLiSt23}. This implies that, up to a certain BER threshold, although there will be discrepancies between the actual and stolen weights, these differences may not lead to a significant drop in accuracy, and the stolen AI model will still achieve the baseline accuracy. However, for higher BER values, accuracy may begin to degrade, which is undesirable. In this case, as explained in Section \ref{subsec:vote}, Alice can retrieve the weights multiple times and use a voting scheme to reduce the BER below a threshold, ensuring that accuracy remains unaffected.

The attack is agnostic to the AI model (e.g., multilayer perceptrons, convolutional neural networks, recurrent neural networks, graph neural networks, spiking neural networks, transformers, etc.) and applicable to any AI hardware accelerator. This is because all accelerators utilize on-chip memory to store model weights, and the HT extracts the weights directly from this memory without interfering with the rest of the architecture—such as the compute units, scheduler, or interconnects—which are typically tailored to the specific AI model type. The attack is also data-agnostic and inference-independent. It is not actively learning the weights by querying the AI model using it as an oracle. 

The success of the attack is determined by the following metrics:

\begin{itemize}
    \item \textit{Accuracy}: The accuracy of the AI model with approximate stolen weights due to the BER.
    \item \textit{Leakage time}: The duration required for Eve to steal the full weight matrix, multiplied by the number of times the matrix is leaked in order to reduce the BER to levels where the baseline accuracy is maintained. Leakage time is influenced by factors such as the model size, data precision, the number of bits the covert channel can carry per frame, and the BER.
    \item \textit{Stealthiness}: The footprint of the HT which should be small to evade detection and the transparency of the covert channel so as to be imperceptible by Alice and Bob.
\end{itemize}

\section{Covert Channel}
\label{sec:covert_channel}

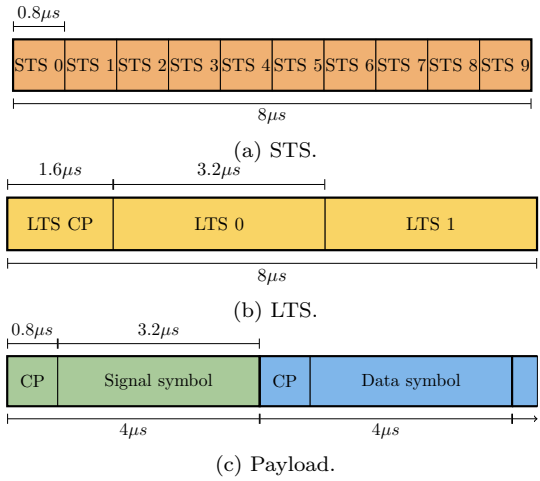
\begin{figure}
    \centering
    \begin{subfigure}{0.8\linewidth}
        \resizebox{\linewidth}{!}{\begin{tikzpicture}
	\begin{pgfonlayer}{nodelayer}
		\node [style=none] (0) at (0, 0) {};
		\node [style=none] (1) at (1, 0) {};
		\node [style=none] (20) at (1, -1) {};
		\node [style=none] (21) at (0, -1) {};
		\node [style=none] (22) at (1, 0) {};
		\node [style=none] (23) at (2, 0) {};
		\node [style=none] (24) at (2, -1) {};
		\node [style=none] (25) at (1, -1) {};
		\node [style=none] (26) at (2, 0) {};
		\node [style=none] (27) at (3, 0) {};
		\node [style=none] (28) at (3, -1) {};
		\node [style=none] (29) at (2, -1) {};
		\node [style=none] (30) at (3, 0) {};
		\node [style=none] (31) at (4, 0) {};
		\node [style=none] (32) at (4, -1) {};
		\node [style=none] (33) at (3, -1) {};
		\node [style=none] (34) at (4, 0) {};
		\node [style=none] (35) at (5, 0) {};
		\node [style=none] (36) at (5, -1) {};
		\node [style=none] (37) at (4, -1) {};
		\node [style=none] (38) at (5, 0) {};
		\node [style=none] (39) at (6, 0) {};
		\node [style=none] (40) at (6, -1) {};
		\node [style=none] (41) at (5, -1) {};
		\node [style=none] (42) at (6, 0) {};
		\node [style=none] (43) at (7, 0) {};
		\node [style=none] (44) at (7, -1) {};
		\node [style=none] (45) at (6, -1) {};
		\node [style=none] (46) at (7, 0) {};
		\node [style=none] (47) at (8, 0) {};
		\node [style=none] (48) at (8, -1) {};
		\node [style=none] (49) at (7, -1) {};
		\node [style=none] (50) at (9, 0) {};
		\node [style=none] (51) at (10, 0) {};
		\node [style=none] (52) at (10, -1) {};
		\node [style=none] (53) at (9, -1) {};
		\node [style=none] (54) at (8, 0) {};
		\node [style=none] (55) at (9, 0) {};
		\node [style=none] (56) at (9, -1) {};
		\node [style=none] (57) at (8, -1) {};
		\node [style=Text] (58) at (0.5, -0.5) {STS 0};
		\node [style=Text] (59) at (1.5, -0.5) {STS 1};
		\node [style=Text] (60) at (2.5, -0.5) {STS 2};
		\node [style=Text] (61) at (3.5, -0.5) {STS 3};
		\node [style=Text] (62) at (4.5, -0.5) {STS 4};
		\node [style=Text] (63) at (5.5, -0.5) {STS 5};
		\node [style=Text] (64) at (6.5, -0.5) {STS 6};
		\node [style=Text] (65) at (7.5, -0.5) {STS 7};
		\node [style=Text] (66) at (8.5, -0.5) {STS 8};
		\node [style=Text] (67) at (9.5, -0.5) {STS 9};
		\node [style=none] (103) at (0, 0.25) {};
		\node [style=none] (104) at (1, 0.25) {};
		\node [style=none] (105) at (0.5, 0.5) {$0.8 \mu s$};
		\node [style=none] (106) at (0, -1.25) {};
		\node [style=none] (107) at (10, -1.25) {};
		\node [style=none] (108) at (5, -1.5) {$8 \mu s$};
	\end{pgfonlayer}
	\begin{pgfonlayer}{edgelayer}
		\draw [style=sts] (0.center)
			 to (1.center)
			 to (20.center)
			 to (21.center)
			 to cycle;
		\draw [style=sts] (22.center)
			 to (23.center)
			 to (24.center)
			 to (25.center)
			 to cycle;
		\draw [style=sts] (26.center)
			 to (27.center)
			 to (28.center)
			 to (29.center)
			 to cycle;
		\draw [style=sts] (30.center)
			 to (31.center)
			 to (32.center)
			 to (33.center)
			 to cycle;
		\draw [style=sts] (34.center)
			 to (35.center)
			 to (36.center)
			 to (37.center)
			 to cycle;
		\draw [style=sts] (38.center)
			 to (39.center)
			 to (40.center)
			 to (41.center)
			 to cycle;
		\draw [style=sts] (42.center)
			 to (43.center)
			 to (44.center)
			 to (45.center)
			 to cycle;
		\draw [style=sts] (46.center)
			 to (47.center)
			 to (48.center)
			 to (49.center)
			 to cycle;
		\draw [style=sts] (50.center)
			 to (51.center)
			 to (52.center)
			 to (53.center)
			 to cycle;
		\draw [style=sts] (54.center)
			 to (55.center)
			 to (56.center)
			 to (57.center)
			 to cycle;
		\draw [style=Thick] (52.center)
			 to (51.center)
			 to (0.center)
			 to (21.center)
			 to cycle;
		\draw [style=double stop] (103.center) to (104.center);
		\draw [style=double stop] (106.center) to (107.center);
	\end{pgfonlayer}
\end{tikzpicture}}
        \caption{STS.}
        \label{fig:sts}
    \end{subfigure}
    \begin{subfigure}{0.8\linewidth}
        \resizebox{\linewidth}{!}{\begin{tikzpicture}
	\begin{pgfonlayer}{nodelayer}
		\node [style=none] (68) at (0, -1) {};
		\node [style=none] (69) at (0, 0) {};
		\node [style=none] (70) at (2, 0) {};
		\node [style=none] (71) at (2, -1) {};
		\node [style=none] (72) at (2, -1) {};
		\node [style=none] (73) at (2, 0) {};
		\node [style=none] (74) at (6, 0) {};
		\node [style=none] (75) at (6, -1) {};
		\node [style=none] (76) at (6, -1) {};
		\node [style=none] (77) at (6, 0) {};
		\node [style=none] (78) at (10, 0) {};
		\node [style=none] (79) at (10, -1) {};
		\node [style=none] (80) at (1, -0.5) {LTS CP};
		\node [style=none] (81) at (4, -0.5) {LTS 0};
		\node [style=none] (82) at (8, -0.5) {LTS 1};
		\node [style=none] (103) at (0, 0.25) {};
		\node [style=none] (104) at (2, 0.25) {};
		\node [style=none] (105) at (1, 0.5) {$1.6 \mu s$};
		\node [style=none] (106) at (2, 0.25) {};
		\node [style=none] (107) at (6, 0.25) {};
		\node [style=none] (108) at (4, 0.5) {$3.2 \mu s$};
		\node [style=none] (109) at (0, -1.25) {};
		\node [style=none] (110) at (10, -1.25) {};
		\node [style=none] (111) at (5, -1.5) {$8\mu s$};
	\end{pgfonlayer}
	\begin{pgfonlayer}{edgelayer}
		\draw [style=lts] (71.center)
			 to (68.center)
			 to (69.center)
			 to (70.center)
			 to cycle;
		\draw [style=lts] (75.center)
			 to (72.center)
			 to (73.center)
			 to (74.center)
			 to cycle;
		\draw [style=lts] (79.center)
			 to (76.center)
			 to (77.center)
			 to (78.center)
			 to cycle;
		\draw [style=double stop] (103.center) to (104.center);
		\draw [style=Thick] (69.center)
			 to (68.center)
			 to (79.center)
			 to (78.center)
			 to cycle;
		\draw [style=double stop] (106.center) to (107.center);
		\draw [style=double stop] (109.center) to (110.center);
	\end{pgfonlayer}
\end{tikzpicture}}
        \caption{LTS.}
        \label{fig:lts}
    \end{subfigure}
    \begin{subfigure}{0.8\linewidth}
        \resizebox{\linewidth}{!}{\begin{tikzpicture}
	\begin{pgfonlayer}{nodelayer}
		\node [style=none] (83) at (0, 0) {};
		\node [style=none] (84) at (1, 0) {};
		\node [style=none] (85) at (1, -1) {};
		\node [style=none] (86) at (0, -1) {};
		\node [style=none] (87) at (1, 0) {};
		\node [style=none] (88) at (5, 0) {};
		\node [style=none] (89) at (5, -1) {};
		\node [style=none] (90) at (1, -1) {};
		\node [style=none] (91) at (0.5, -0.5) {CP};
		\node [style=none] (92) at (3, -0.5) {Signal symbol};
		\node [style=none] (93) at (5, 0) {};
		\node [style=none] (94) at (6, 0) {};
		\node [style=none] (95) at (6, -1) {};
		\node [style=none] (96) at (5, -1) {};
		\node [style=none] (97) at (6, 0) {};
		\node [style=none] (98) at (10, 0) {};
		\node [style=none] (99) at (10, -1) {};
		\node [style=none] (100) at (6, -1) {};
		\node [style=none] (101) at (5.5, -0.5) {CP};
		\node [style=none] (102) at (8, -0.5) {Data symbol};
		\node [style=none] (103) at (0, 0.25) {};
		\node [style=none] (104) at (1, 0.25) {};
		\node [style=none] (105) at (0.5, 0.5) {$0.8 \mu s$};
		\node [style=none] (106) at (0, -1.25) {};
		\node [style=none] (107) at (5, -1.25) {};
		\node [style=none] (108) at (2.5, -1.5) {$4\mu s$};
		\node [style=none] (109) at (1, 0.25) {};
		\node [style=none] (110) at (5, 0.25) {};
		\node [style=none] (111) at (3, 0.5) {$3.2 \mu s$};
		\node [style=none] (112) at (10, -1) {};
		\node [style=none] (113) at (10, 0) {};
		\node [style=none] (114) at (10.5, 0) {};
		\node [style=none] (115) at (10.5, -1) {};
		\node [style=none] (116) at (5, -1.25) {};
		\node [style=none] (117) at (10, -1.25) {};
		\node [style=none] (118) at (7.5, -1.5) {$4\mu s$};
		\node [style=none] (119) at (10.5, -1.25) {};
	\end{pgfonlayer}
	\begin{pgfonlayer}{edgelayer}
		\draw [style=signal] (84.center)
			 to (83.center)
			 to (86.center)
			 to (85.center)
			 to cycle;
		\draw [style=signal] (88.center)
			 to (87.center)
			 to (90.center)
			 to (89.center)
			 to cycle;
		\draw [style=data] (94.center)
			 to (93.center)
			 to (96.center)
			 to (95.center)
			 to cycle;
		\draw [style=data] (100.center)
			 to (99.center)
			 to (98.center)
			 to (97.center)
			 to cycle;
		\draw [style=double stop] (103.center) to (104.center);
		\draw [style=Thick] (83.center)
			 to (88.center)
			 to (89.center)
			 to (86.center)
			 to cycle;
		\draw [style=Thick] (93.center)
			 to (98.center)
			 to (99.center)
			 to (96.center)
			 to cycle;
		\draw [style=double stop] (106.center) to (107.center);
		\draw [style=double stop] (109.center) to (110.center);
		\draw [style=data] (113.center)
			 to (112.center)
			 to (115.center)
			 to (114.center)
			 to cycle;
		\draw [style=Thick] (115.center) to (112.center);
		\draw [style=Thick] (112.center) to (113.center);
		\draw [style=Thick] (113.center) to (114.center);
		\draw [style=double stop] (116.center) to (117.center);
		\draw [style=stop arrow] (117.center) to (119.center);
	\end{pgfonlayer}
\end{tikzpicture}}
        \caption{Payload.}
        \label{fig:payload}
    \end{subfigure}
    \caption{Frame format of an OFDM IEEE 802.11 transmission.}\vspace{-0.0cm}
    \label{fig:wifi}
\end{figure}

We use a state-of-the-art covert channel proposed in \cite{DiRiAbSt22}, but with a fundamental different hardware implementation. Herein, we describe the theory of the covert channel, while in Section \ref{subsec:HT-Alice} we discuss the hardware implementation in Alice.

We consider the \ac{OFDM}-based IEEE 802.11 protocol for \ac{WLAN} commonly known as Wi-Fi \cite{ieee80211}. The frame format for transmission is organized into 3 parts, as shown in \cref{fig:wifi}. The first part is the preamble composed of two fields, namely the \ac{STS} and the Long Training Sequence (LTS), as shown in \cref{fig:sts,fig:lts}. The STS is used by the receiver for detecting the start of the frame, for automatic gain control, for coarse timing synchronization, and for coarse frequency offset estimation. The LTS is used for fine timing synchronization, fine frequency offset estimation, and helps the receiver in channel estimation. The second and third parts, shown in \cref{fig:payload}, are, respectively, the signal, which carries control information to help the receiver decode the data (i.e., rate, modulation, length, etc.), and the payload, i.e., the actual data being transmitted.

\begin{table}
    \centering
    \caption{Frequency domain definition of the \acs{STS}.}
    \begin{tabular}{ccc}
        Index & I & Q \\
        \hline
        -24, -16, -4, 12, 16, 20, 24 & $\sqrt{\frac{13}{6}}$ & $\sqrt{\frac{13}{6}}$\\
        -20, -12, -8, 4, 8 & $-\sqrt{\frac{13}{6}}$ & $-\sqrt{\frac{13}{6}}$\\
    \end{tabular}\vspace{-0.0cm}
    \label{tab:subcar-val}
\end{table}

The covert channel is hidden in the \ac{STS} of the preamble without affecting the receiver's capacity to synchronize the packet correctly. More specifically, the \ac{STS} defined in the frequency domain contains a single \ac{OFDM} symbol or 64 samples. Each sample is a subcarrier or frequency bin. In \ac{OFDM}, samples are represented as complex numbers having a real (I) and an imaginary (Q) part (In-phase and Quadrature components). Out of the 64 subcarriers, 12 have a non-zero amplitude. The indexes and values of these non-zero subcarriers are shown in Table \ref{tab:subcar-val}. The covert channel hides one byte of information in 8 of the non-zero subcarriers of the STS, called corrupted subcarriers, i.e., one bit per subcarrier. More specifically, the subcarrier magnitude is multiplied by a factor of $1-\alpha$, $0<\alpha<1$, if the leaked bit is 1, whereas the value is unchanged if the leaked bit is 0. The 8 corrupted subcarriers are arbitrarily selected to be those with indexes $k= \{-24,-20, -16, -8, 4, 8, 16, 24\}$ and remain the same across frames. 

The time domain STS is obtained by performing an inverse Fast Fourier Transform (IFFT) of the frequency domain definition. It contains 4 repetitions of 16 complex-valued IQ samples with duration $0.8\mu s$ each. We refer to one repetition as short \ac{STS}. The complete time domain \ac{STS}, shown in \cref{fig:sts}, consists of two and a half time domain \ac{STS} symbols, i.e., 10 short \ac{STS}, named STS0 to STS9 in \cref{fig:sts}.

As we will see, the parameter $\alpha$ governs both the transparency of the covert channel and Eve's ability to recover the leaked information with low BER. Specifically, a smaller $\alpha$ results in a more stealthy covert channel, while a larger $\alpha$ leads to a lower BER for Eve at a given SNR.

This covert channel, although demonstrated for Wi-Fi, it virtually applies to any communication protocol whose synchronization process is based on preamble correlation, such as Bluetooth and ZigBee.

\section{Hardware Implementation}
\label{sec:implementation}

\subsection{Hardware platform}

The hardware demonstrator employs the Software Defined Radio (SDR) bladeRF 2.0 micro xA9 board from Nuand \cite{bladeRF}. It is composed of a Cypress FX3 microcontroller, a fully programmable Cyclone V Field Programmable Gate Array (FPGA) from Intel, and an RF transceiver AD9361 from Analog Devices. 

Alice and Eve are implemented using two separate bladeRF boards. The HT enabling the covert channel in Alice is embedded into the PHY layer. The PHY layer prepares the frame with the appended preamble for transmission. To implement the PHY layer, we use the open source \textit{bladeRF-wiphy} project \cite{bladeRF-wiphy} written in VHDL and we modify it accordingly to embed the HT. 

In this initial hardware prototype, we omit the AI hardware accelerator from Alice due to the limited FPGA resources on the bladeRF board. Instead, the weight matrix of the AI models is supplied externally.  However, this is without loss of generality since, as mentioned, the attack is agnostic to the AI hardware accelerator requiring only access to the weights in memory.

\subsection{Alice implementation}\label{subsec:HT-Alice}

In nominal HT-free operation, \ac{STS} has a fixed value. The HT modulates this value to leak gradually the weight matrix over several transmitted frames. In the \textit{bladeRF-wiphy} project, only a single time-domain short \ac{STS} is stored, which thereafter is repeated 10 times to create the complete \ac{STS} sequence to be prepended to each transmitted frame, as shown in \cref{fig:sts}. This is to avoid having to compute the IFFT of the frequency-domain representation every time a frame is being generated.

\begin{figure}
    \centering
    \resizebox{0.7\linewidth}{!}{\includesvg[width=0.8\linewidth]{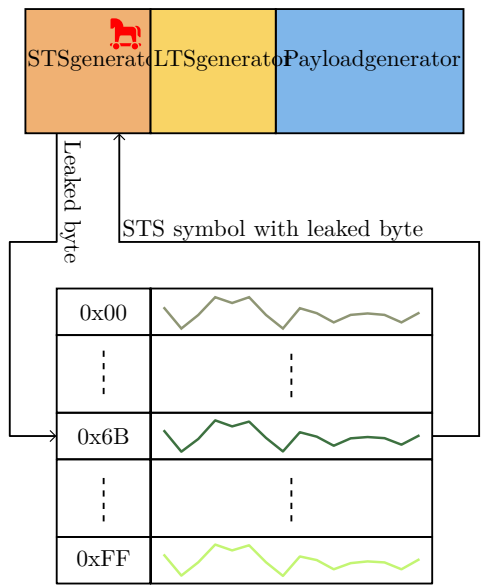}}
    \caption{Lookup table inside the \ac{STS} generator block.}\vspace{-0.0cm}
    \label{fig:sts-table}
\end{figure}

One byte of the weight matrix is leaked per transmitted frame. In \cite{DiRiAbSt22}, the \ac{STS} is modulated with the leaked byte in the frequency domain. In contrast, herein we propose a more efficient implementation in the time domain. More specifically, given that one byte is leaked, there are $2^8=256$ possible values of \ac{STS} in each transmission. We generate a lookup table that contains the $256$ pre-calculated values of the short \ac{STS} in the time domain. Fig. \ref{fig:sts-table} illustrates this table, where each row represents the leaked byte in hexadecimal along with its corresponding real (I) time-domain \ac{STS} waveform using $\alpha=0.15\%$. Notice that for each leaked byte the waveforms have subtle differences that are hardly visible in \cref{fig:sts-table}.
This table is stored in a memory within the \ac{STS} generator. The value of $\alpha$ needs to be pre-defined since for a different $\alpha$ value a different lookup table needs to be computed. 

The HT mechanism reads the weight matrix and fetches 1 byte to the \ac{STS} generator to be leaked per transmitted frame, until the complete matrix is leaked. Then, the short \ac{STS} corresponding to the byte value is selected from the lookup table, and is repeated 10 times to generate the \ac{STS} carrying the covert channel information, which is then appended to the transmitted frame.

In short, only the \ac{STS} generator block is modified in the PHY layer to add the lookup table memory, with the addition of the connection from the weight memory to the \ac{STS} generator block to drive 1 byte of memory to be leaked per transmitted frame. The byte value defines the lookup table row to be used and the corresponding \ac{STS} is being prepended to the transmitted frame. The rest of the transmitted frame, i.e., LTS, signal, and payload, are left unchanged.

\subsection{Footprint of HT in Alice}\label{subsec:HT_overhead}

\begin{table}
    \centering
    \caption{HT overhead.}
    \begin{tabular}{c|cc}
        & HT-free & HT-infected \\
        \hline
        Logic utilization & 35,751 (31 \%) & 36,104 (32 \%) \\
        Total registers & 56015 & 55916 \\
        Total pins & 173 (77 \%)& 173 (77 \%)\\
        Total virtual pins & 0& 0\\
        Total block memory bits & 2,137,500 (17 \%)& 2,137,500 (17 \%)\\
        Total RAM Blocks & 324 (27 \%)& 324 (27 \%)\\
        Total DSP Blocks & 100 (29 \%)& 100 (29 \%)\\
        Total HSSI RX PCSs & 0&0\\
        Total HSSI TX PCSs & 0&0\\
        Total PLLs & 4 (50 \%)& 4 (50 \%)\\
        Total DLLs & 0&0\\
    \end{tabular}\vspace{-0.0cm}
    \label{tab:overhead}
\end{table}

Table \ref{tab:overhead} shows the registers and logic modules overhead of the HT-infected PHY layer of the RF transmitter of Alice with respect to the HT-free design after synthesizing the two designs using Quartus II 16.0 from Intel. There is 1\% more logic utilization and a smaller number of registers, which might as well be due to the inherent optimizations performed by the synthesis tool. Thus, the area overhead is minimal or even negligible. Similarly, we did not observe any appreciable power consumption overhead. Thus, we conclude that the HT has a minimum footprint making it extremely stealthy.







\subsection{Transparency of covert channel}\label{subsec:transparency_covert_channel}

\begin{figure}
    \centering \includesvg[width=0.9\columnwidth]{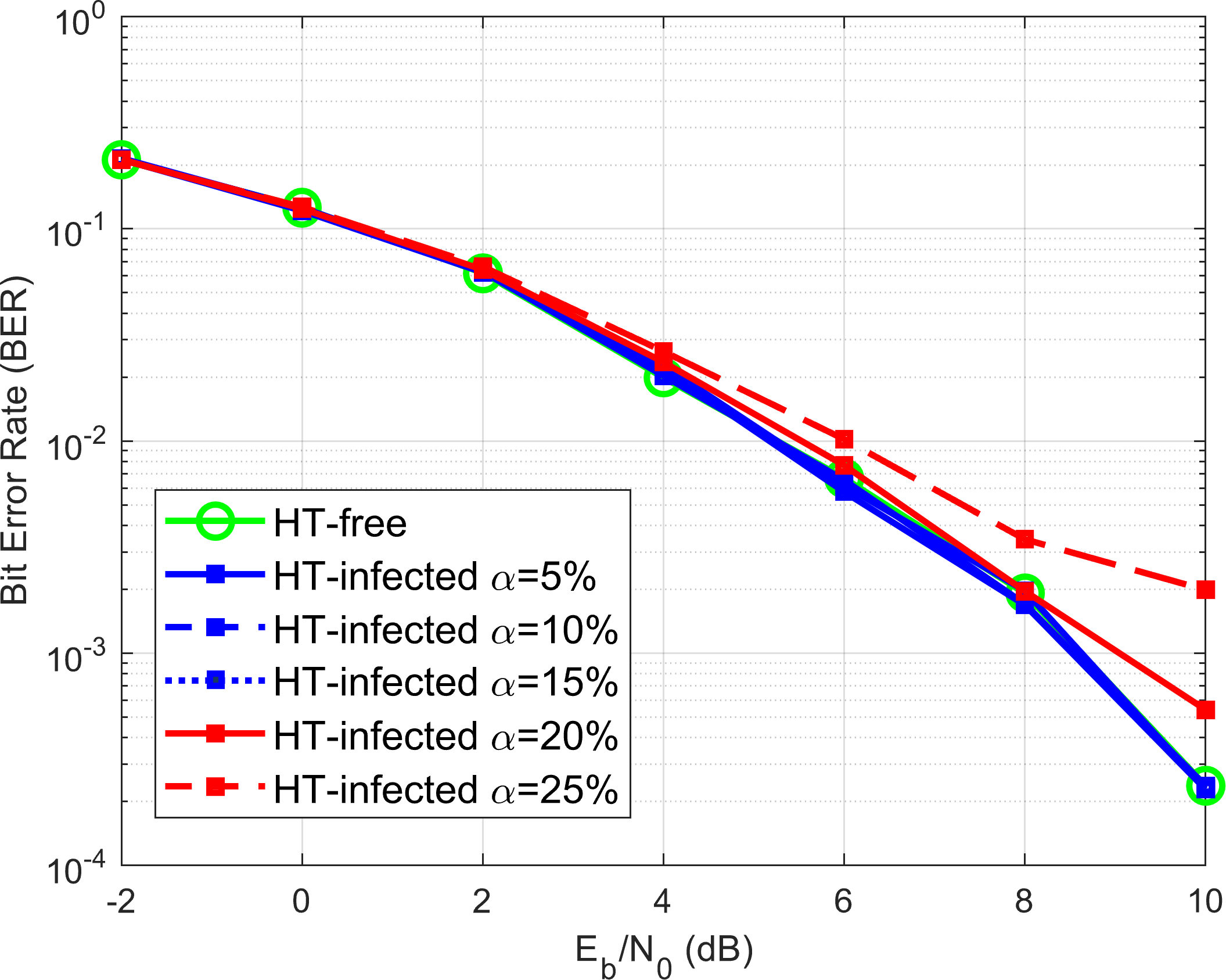}
    \caption{Measured BER using a HT-free and an HT-infected Alice device for different $\alpha$ values.}\vspace{-0.0cm}
    \label{fig:payload-ber}
\end{figure}

Fig. \ref{fig:payload-ber} shows the BER as a function of SNR for the HT-free and HT-infected Alice device as a function of $\alpha$. As it can be seen, for $\alpha \leq 15\%$, the covert channel in the preamble has no performance penalty on the communication as we observe the same BER, essentially making the covert channel totally transparent and undetectable by Bob. In contrast, for $\alpha \geq 20\%$, we observe a deterioration of BER of the regular receiver. The highest possible value of $\alpha$ should be used since it increases the strength of the covert channel and lowers the BER of the reconstructed weight matrix by Alice. Therefore, we conclude that the best choice of $\alpha$ is 15\%.

\subsection{Throughput of Covert channel}\label{subsec:covert_channel_throughput}

The throughput of the covert channel measured in \ac{Bps} depends on the channel occupation between Alice and the nominal receiver Bob. The payload that Alice transmits to Bob is split into several frames and each frame transmission cycle is composed of the following steps: (a) before attempting a transmission Alice waits for the channel to be idle for a minimum time, called Distributed Coordination Function (DCF) Interframe Space (DIFS); (b) Alice transmits the frame to Bob; (c) Bob gets a priority time window called Short Interframe Space (SIFS) shorter than DIFS to respond before other devices might think the channel is free after DIFS and start transmitting, causing a collision; (d) Bob sends an acknowledgment (ACK) to confirm successful reception. This cycle is repeated until the complete payload is transmitted. 

A frame consists of a fixed-length part and a variable-length part. The fixed-length part lasts for $20\mu s$ and is composed of the preamble and the signal, as shown in \cref{fig:wifi}. The variable-length part is composed of a variable number of payload data symbols, with a maximum size of 1500 bytes, lasting for $\approx224\mu s$ at 54 Mbps. 
The DIFS lasts for $\approx28\mu s$, the SIFS for $\approx10\mu s$, and an ACK frame
is transmitted in $\approx40\mu s$. Therefore, in this scenario, a complete cycle lasts $\approx 322 \mu s$ and within this time 1 byte is leaked, resulting in a throughput of about $T=3105$ \ac{Bps}. This throughput increases when the payload part of the transmitted frames becomes shorter and when the channel traffic is reduced or the channel is idle.

\subsection{Leakage time}\label{sec:leakage_time}

The time $T_{leak}$ required to leak the parameters of the AI model is given by 

\begin{equation}
    T_{leak} = \frac{N_f}{T},
\end{equation}

\noindent where $N_f$ is the total number of frames that need to be transmitted to leak all parameters and $T$ is the throughput of the covert channel.

\subsection{Eve implementation}\label{subsec:receiver}

In a typical receiver, e.g. Bob, the \ac{STS} is used to detect incoming frames. Once it has served its purpose, it is discarded and not processed further.

In contrast, for an eavesdropper like Eve, the \ac{STS} contains leaked information and must therefore be stored and analyzed to extract this data. This necessitates a specialized receiver capable of such processing. To this end, we implemented Eve using the open-source \textit{bladeRF-wiphy} framework and deployed it on a second bladeRF device to monitor Alice’s transmissions.

Specifically, the additional operations performed by Eve, beyond those of the nominal receiver Bob, are as follows. First, the full preamble is used to perform phase and frequency offset correction on the \ac{STS}, applying the same techniques typically used for the payload. An FFT is then applied to the resulting \ac{STS} to convert it from the time domain to the frequency domain, allowing Eve to extract the amplitudes of the 8 corrupted subcarriers encoding the leaked byte and infer the corresponding bits.

\begin{figure}
    \centering
    \resizebox{.9\linewidth}{!}{\input{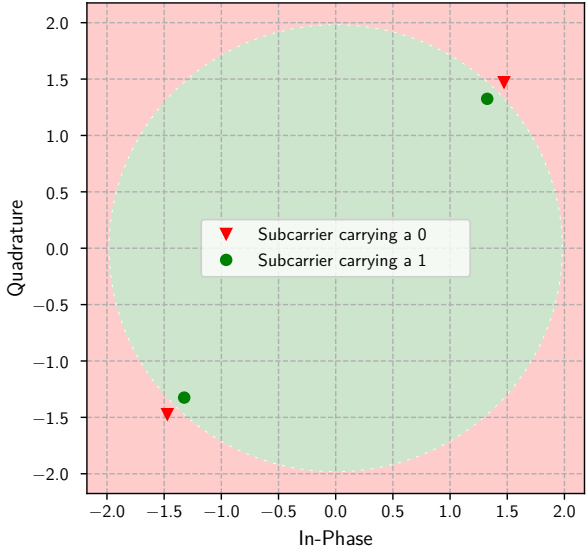}}
    \caption{Threshold operation for reliable extraction of the leaked bit.}
    \label{fig:eve-threshold}
\end{figure}

In our implementation, the remaining 4 non-zero subcarriers which do not carry leaked bits are utilized as reference  to improve the reliability of bit extraction. This design choice explains why we opted to leak only 1 byte per transmission, even though leaking 12 bits was possible and would have increased throughput. Specifically, the magnitudes of these 4 reference subcarriers are averaged to form a reference magnitude. This value is then scaled by a threshold factor, empirically set to $1-\frac{\alpha}{2}$, to define the threshold magnitude. Each corrupted subcarrier is then compared against this threshold: if its magnitude exceeds the threshold, the corresponding bit is interpreted as 1; otherwise, it is interpreted as 0. An illustration of this decision process is shown in \cref{fig:eve-threshold} using a constellation diagram. The green-shaded area represents subcarriers corresponding to leaked bit 1, with the boundary between the green and red regions indicating the threshold magnitude.

\subsection{Leakage repetition and voting}\label{subsec:vote}

\begin{table}[t]
    \centering
    \caption{Example of the voting mechanism by Eve.}
    \begin{tabular}{c|c c c c c cc c}
        Decimal & \multicolumn{8}{c}{MSB \Arrow{2.5cm} LSB}\\
        \hline
        \hline
        160&1 & 0 &1&0&0&0&0&0 \\
        163&1 & 0 &1&0&0&0&1&1 \\
        160&1 & 0 &1&0&0&0&0&0 \\
        \hline
        Number of ones &\color{green} 3 & \color{red}0&\color{green}3&\color{red}0&\color{red}0&\color{red}0&\color{red}1&\color{red}1 \\
        \hline
        160&1 & 0 &1&0&0&0&0&0 \\
    \end{tabular}
    \label{tab:voting}
\end{table}

Due to the BER in the communication channel, Eve may only recover an approximate version of the weight matrix. If the resulting model accuracy is insufficient, she can moderate the effect of the BER and recover the original baseline accuracy by receiving multiple broadcasts of the weight matrix and applying a voting scheme. Specifically, Eve collects an odd number of noisy copies of each weight, where some bits may be flipped due to transmission errors. The voting process is performed bit-wise, selecting the most frequent bit value (0 or 1) across the received copies. An illustrative example is provided in Table \ref{tab:voting}, where a single weight is represented as a 1-byte value and received three times. In this case, Alice transmits a weight with decimal value 160. Eve receives this byte correctly twice, while in one instance it is corrupted to 163. Using the voting scheme, Eve correctly reconstructs the original byte value of 160, effectively reducing the BER for this weight to zero.

\section{Case Studies}
\label{sec:case_studies}

\begin{table}[t]
    \caption{Case studies and leakage time.}
    \label{tab:leak-time}
    \centering
    \begin{tabular}{|c|c|c|c|}
    \hline
         & Data & Memory & \\
         Model & precision  & usage &$T_{leak}$\\
         & (Bytes) & (Bytes) & \\
         \hline
         \hline
         LeNet5 & 4 & 246,824 &  79s \\
         \hline
         Quantized LeNet5 & 1 & 61,470 & 20s \\
         \hline
         MobileNetV3-Large & 4 & 21,932,128 & 1h58m \\
         \hline
        IBM DVS128 Gesture SNN & 4 & 16,954,240 & 1h31m \\
         \hline
         YOLO11n & 4 & 10,464,992 & 56m16s \\
         \hline
    \end{tabular}
\end{table}

Table \ref{tab:leak-time} summarizes the different AI models used as case studies, indicating their data precision and memory footprint for parameter storage. Specifically, the AI models are:

\subsubsection{LeNet5}

LeNet5 \cite{LBBH98} is a convolutional neural network (CNN) designed for the MNIST dataset \cite{mnist}, which consists of 70,000 grayscale images of handwritten digits.
Of these, 60,000 images are used for training and 10,000 for testing. We trained two versions of LeNet5 using PyTorch \cite{PyTorch2}, one that uses single precision floating-point (32-bit) and a second 8-bit integer quantized version. For the quantized version, we performed quantization-aware training (QAT) using the Brevitas open-source PyTorch framework \cite{brevitas}.


\subsubsection{MobileNetV3-Large}

MobileNetV3-Large \cite{mobileNetV3} is a lightweight CNN architecture designed for mobile and edge devices, developed by Google. It is trained using PyTorch \cite{PyTorch2} on the ImageNet dataset \cite{imagenet}, which consists of 1,2 million training images and 50,000 validation images across 1,000 classes.

\subsubsection{IBM's DVS128 Gesture SNN}

We trained a SNN on the IBM's DVS128 gesture dataset \cite{ATBM17}, consisting of 29 individuals performing 11 hand and arm gestures in front of a dynamic vision sensor, such as hand waving and air guitar, under 3 different lighting conditions. Training was conducted using the Spike LAYer Error Reassignment (SLAYER) framework \cite{shor18}, which enables backpropagation tailored for SNNs. The dataset comprises 1,342 spiking-format samples, with data from the first 23 individuals used for training and the remaining 6 individuals reserved for testing.

\subsubsection{YOLO11n}

\acf{YOLO} \cite{yolo} is a family of real-time object detection algorithms that frame object detection as a single regression problem, directly predicting bounding boxes and class probabilities from full images in one evaluation. The version of \ac{YOLO} used for this work is YOLO11n \cite{ultralytics2024yolov11}, trained on the \ac{COCO} dataset \cite{COCO}, which is made up of 330,000 images, with more than 200,000 labeled. There are 80 categories of objects to recognize. The inference was performed in PyTorch \cite{PyTorch2} using a subset of \ac{COCO} named COCO8 \cite{ultralytics2023coco8}. It comprises the first 8 images from the COCO training set, with 4 images designated for training and 4 for validation. Despite its small size, COCO8 offers sufficient diversity to serve as a practical dataset for experimenting with YOLO.

\section{Experimental Results}
\label{sec:results}

\subsection{Leakage time}

The fourth column of \cref{tab:leak-time} presents the time needed to leak the model once, assuming a throughput of $T=3105$ \ac{Bps}, as discussed in Section \ref{subsec:covert_channel_throughput}. This leakage time ranges from a few seconds for the smaller LeNet-5 model to approximately two hours for the largest MobileNetV3-Large model.

\subsection{BER of covert channel}

\begin{figure}
    \centering
    \resizebox{.9\linewidth}{!}{\input{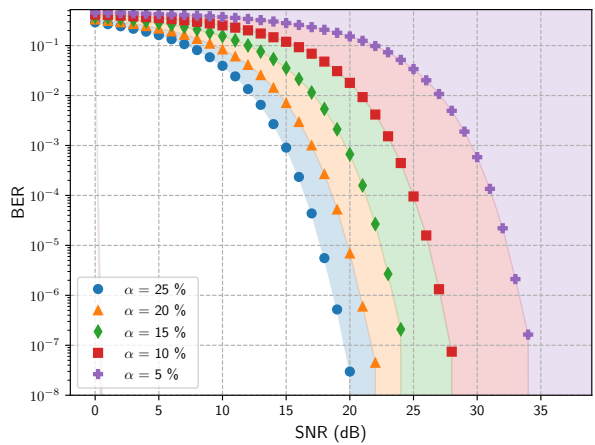}}
    \caption{Effect of $\alpha$ and SNR on the \ac{BER} of Eve.}\vspace{-0.0cm}
    \label{fig:ber-alpha}
\end{figure}

The \ac{BER} of the covert channel is influenced by Eve's SNR—that is, the ability of Eve to distinguish the intended signal from background noise—as well as the $\alpha$ parameter in the HT mechanism. Fig. \ref{fig:ber-alpha}  presents hardware measurement results illustrating how both SNR and $\alpha$ impact the BER. As expected, higher SNR and larger $\alpha$ values lead to lower BER. In Section \ref{subsec:transparency_covert_channel}, Fig. \ref{fig:payload-ber} showed that setting $\alpha \leq 15\%$ is necessary to keep the covert channel imperceptible to Bob. However, in Fig. \ref{fig:ber-alpha} we include larger $\alpha$ values to explore their effect on BER. Additionally, achieving reliable high-speed Wi-Fi typically requires SNR $>20$ dB. Based on these observations, we define two scenarios:

\begin{itemize}
    \item \textit{Scenario 1}: High \ac{SNR}$=30$ dB and $\alpha=15\%$. This is the most favorable scenario for Eve as the BER is lower than $10^{-8}$, as it can be seen from Fig. \ref{fig:ber-alpha}. 
    \item \textit{Scenario 2}: Minimum \ac{SNR}$=20$ dB and $\alpha=15\%$. From Fig. \ref{fig:ber-alpha}, the BER is $\approx 7\cdot10^{-4}$ indicating that Eve reconstructs the AI model with significant weight perturbations.
\end{itemize}

\subsection{Leaking AI models}

The four models in Section \ref{sec:case_studies} were leaked using our hardware implementation of the covert channel, under channel conditions that produced varying BER. For each experiment, a full inference was performed to obtain the accuracy of the model with stolen weights. Fig. \ref{fig:acc-ber} shows the computed accuracy as a function of BER across all models. For MobileNetV3-Large, we report both the Top-1 and Top-5 accuracies. For YOLO11, we use the mAP@50 metric to evaluate the accuracy. From Fig. \ref{fig:acc-ber}, we make the following observations:

\begin{figure}
    \centering
    \resizebox{.9\linewidth}{!}{\input{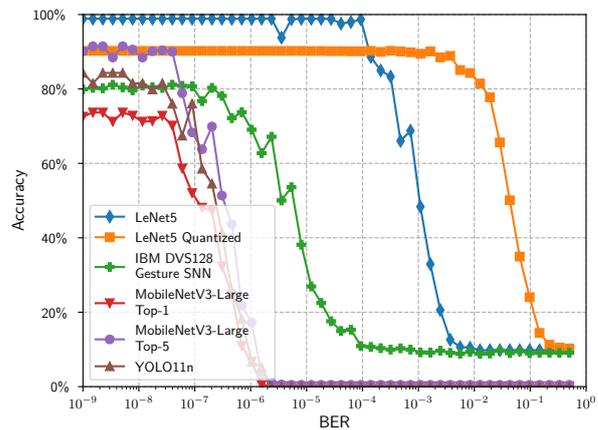}}
    \caption{Accuracy of the different models with stolen parameters as a function of BER.}\vspace{-0.0cm}
    \label{fig:acc-ber}
\end{figure}

\begin{itemize}
    \item As expected, the accuracy drops as the BER increases.
    \item The models exhibit varying levels of resilience to BER. In general, when considering the model sizes presented in \cref{tab:leak-time} (i.e., their memory usage), we observe that larger models tend to experience a drop in accuracy at lower BER values. For instance, LeNet-5, being the smallest model, is significantly more robust to BER—its accuracy only begins to degrade at BER = $10^{-4}$ for the 32-bit floating-point version, and at BER $>10^{-3}$ for the quantized version. In contrast, MobileNetV3-Large, the largest model in the study, starts experiencing accuracy degradation at BER values below $10^{-7}$. The only exception to this trend is the IBM DVS128 Gesture SNN. Despite being significantly larger than YOLO11, it demonstrates greater robustness to BER. However, this comparison involves an SNN versus a level-based artificial neural network (ANN), and while the results suggest that SNNs may exhibit higher resilience than ANNs, a more comprehensive analysis is required to draw general conclusions.
    \item The 8-bit quantized version of LeNet-5 exhibits greater robustness compared to its 32-bit counterpart. This observation, consistent with findings in bit-flip fault injection studies in ANNs \cite{SuLiSt23}, can be attributed to the high sensitivity of floating-point representations. Specifically, a single bit-flip in the exponent of a 32-bit float can lead to a drastic alteration in the weight value, which in turn destabilizes the model and significantly impacts its accuracy.

\end{itemize}

\subsection{Leakage repetition and voting}

\begin{figure}
    \centering
    \resizebox{.9\linewidth}{!}{\input{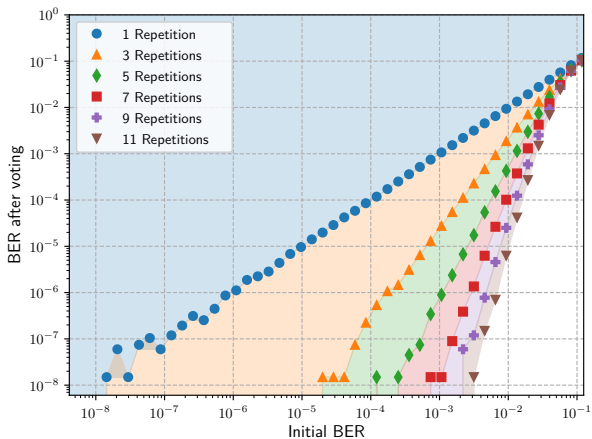}}
    \caption{Improvement in BER with leakage repetitions and voting.}\vspace{-0.0cm}
    \label{fig:ber-reps}
\end{figure}

Referring back to the two scenarios derived from Fig. \ref{fig:ber-alpha}, in the favorable scenario 1 where BER $<10^{-8}$, the results in Fig. \ref{fig:acc-ber} indicate that all models can be reliably leaked in a single broadcast, maintaining their baseline accuracy. In contrast, under the more challenging conditions of scenario 2, where BER $\approx7\cdot10^{-4}$, all models except the quantized version of LeNet-5 experience a sharp drop in accuracy, thus necessitating multiple broadcasts and a voting scheme, as discussed in \cref{subsec:vote}. 

\Cref{fig:ber-reps} illustrates the reduction in \ac{BER} achieved through multiple repetitions of the leakage process followed by majority voting. As shown, the BER decreases significantly as the number of repetitions increases. In scenario 2, performing 3 repetitions reduces the BER to below $10^{-5}$—sufficiently low to leak an approximate 32-bit floating-point LeNet-5 model that maintains baseline accuracy. With 9 repetitions, the BER drops below $10^{-8}$, enabling accurate leakage of all models. The total leakage time is computed from \cref{tab:leak-time} as $r \cdot T_{leak}$, where $r$ is the number of repetitions. If Alice communicates wirelessly with other devices for a duration shorter than $r \cdot T_{leak}$, the leakage proceeds in batches, pausing until  communication is resumed.

By referencing \cref{fig:ber-alpha,fig:acc-ber,fig:ber-reps}, we can determine the number of repetitions needed to leak a model while maintaining baseline accuracy, given specific values for SNR and $\alpha$. First, from \cref{fig:ber-alpha}, we calculate the resulting BER. Then, using \cref{fig:acc-ber}, we identify the minimum BER required to achieve baseline accuracy, and finally, from \cref{fig:ber-reps}, we determine the number of repetitions necessary to reach the required BER.

\section{Countermeasures}
\label{sec:defenses}

According to our threat model, the victim possesses a HT-infected leaking device and has no access to the original design files. As a result, viable countermeasures are limited to post-silicon approaches, focusing either on detecting the HT itself or identifying the covert channel during run-time.

\subsection{Detection of HT}

A traditional approach is reverse engineering, which entails de-packaging the chip, removing its layers, and imaging it to reconstruct the layout and functionality \cite{LWUSEDGRKG19}. However, as shown in Section \ref{subsec:HT_overhead}, the HT has an extremely small footprint, making it difficult to detect through visual inspection. Moreover, reverse engineering is destructive, time-intensive, and costly. Compounding the challenge, the absence of a golden reference design significantly hinders the reliability and effectiveness of detecting malicious modifications.

Logic testing is another widely used method to detect the presence of a HT in a design \cite{CWPPB09}. This approach utilizes a dedicated automatic test pattern generation (ATPG) tool to create test patterns that target rare or infrequently activated paths, as HTs typically activate under uncommon conditions to evade detection. However, in our threat model, such a tool requires access to a gate-level hardware model, which is not available to the defender. Additionally, in our implementation, the HT is always active, making activation conditions irrelevant.

A third method is statistical side-channel fingerprinting (SSCF) \cite{ABKRS07}, which involves collecting chip measurements such as power supply traces, EM emissions, timing information, and temperature. The goal is to distinguish between HT-infected and HT-free chips within this measurement space. The boundary between these two can be defined using a one-class classifier trained on HT-free chip instances. However, this approach requires a golden chip or at least a trusted hardware model, making it incompatible with our threat model. Additionally, the minimal footprint of the proposed HT makes it difficult to differentiate from noise and normal variations.

A fourth approach involves information flow tracking (IFT) methods, which monitor the propagation of sensitive data to ensure it does not reach unauthorized areas in the design \cite{JGDBM17}. In our case, IFT could be used to detect the connection between the weight memory and the preamble generation block. However, IFT is not applicable in our threat model, as it requires access to the Register-Transfer Level (RTL) of the design, which the victim does not have.

\subsection{Identifying the covert channel}

Several works that propose mechanisms for creating covert channels \cite{KKCR13,Dutta13, ClScHo15,HiFr10,DiRiAbSt22, LJNM17,SHANM20, CBOBO18,SRDJWA-SDMIC19} 
also suggest detection defenses using chip testing or during run-time. These defenses range from basic measurements, such as calculating BER, examining compliance with spectral mask specifications, and analyzing IQ constellation diagrams, to more advanced techniques like SSCF \cite{LJNM17} and Adaptive Channel Estimation (ACE) \cite{SHANM20}. The ACE defense takes advantage of the slow-fading characteristics of indoor communication channels to differentiate between channel impairments and HT activity. However, in \cite{DiRiAbSt22}, it is experimentally demonstrated that the covert channel used in our implementation bypasses all known defenses. For example, as shown in Fig. \ref{fig:payload-ber}, we observe that the BER remains identical for both HT-infected and HT-free devices.

In \cite{D-RAAS24}, an AI-based defense mechanism is introduced, which trains a CNN to identify covert channels using IQ samples from transmitted frames encoded into images. An open-source dataset was provided to support this training, containing hardware measurements collected from an HT-infected leaking device featuring various HT implementations. This approach has proven effective in detecting major types of covert channels, including the one in \cite{DiRiAbSt22}. However, if an attacker employs a novel covert channel strategy or one not represented in the dataset, the AI model’s detection will become unreliable or produce misleading results.

A natural defense involves implementing Eve to analyze the preamble. However, designing and fabricating such a specialized receiver exceeds the defender’s capabilities. Moreover, since a novel covert channel could be used, the defender, lacking knowledge of the specific implementation, has no effective means of detecting it.

\section{Conclusion}
\label{sec:conclusion}

We introduced a novel AI model parameter-stealing attack targeting devices that perform AI inference and communicate wirelessly. The attack leverages a covert channel embedded within the wireless transmissions to exfiltrate model parameters without raising suspicion on the victim device. This covert channel is enabled by a minimal-footprint HT. The attack is generic—independent of both the AI hardware accelerator and the AI model—and was validated in hardware using various AI models under different channel conditions. For high SNR scenarios, the stolen model achieves baseline accuracy. In low SNR conditions, repeating the leakage process a few times and applying a voting scheme effectively reduces the BER to a level sufficient for baseline accuracy. Our results demonstrate that even large AI models running on edge devices can be successfully leaked within a few hours.


\begin{acronym}
    \acro{SNR}{Signal to Noise Ratio}
    \acro{IP}{Intellectual Property}
    \acro{AI}{Artificial Intelligence}
    \acro{NN}{Neural Network}
    \acro{OFDM}{Orthogonal Frequency-Division Multiplexing}
    \acro{STS}{Short Training Sequence}
    \acro{LTS}{Long Training Sequence}
    \acro{HT}{Hardware Trojan}
    \acro{BER}{Bit Error Rate}
    \acro{MLP}{MultiLayer Perceptron}
    \acro{IEEE}{Institute of Electrical and Electronics Engineers}
    \acro{SoC}{System on a Chip}
    \acro{fps}{frames per second}
    \acro{RF}{Radio Frequency}
    \acro{WLAN}{Wireless Local Area Network}
    \acro{CFO}{Carrier Frequency Offset}
    \acro{SNN}{Spiking Neural Network}
    \acro{YOLO}{You Only Look Once}
    \acro{COCO}{Common Objects in COntext}
    \acro{Bps}{bytes per seconds}
  \acro{ASIC}{Application-Specific Integrated Circuit}
  \acro{C. elegans}{Caenorhabditis elegans}
  \acro{CIFG}{Coupled Input-Forget Gate}
  \acro{CMOS}{Complementary metal-oxide-semiconductor}
  \acro{CPU}{Central Processing Unit}
  \acro{FCM}{Forward Crawling Motion}
  \acro{FGR}{Full Gate Recurrence}
  \acro{FPGA}{Field Programmable Gate Array}
  \acro{GPT}{Generative Pre-trained Transformer}
  \acro{GPU}{Graphics Processing Unit}
  \acro{GRU}{Gated Recurrent Unit}
  \acro{LIDM}{Linear Ion Drift Model}
  \acro{LLaMA}{Large Language Model Meta AI}
  \acro{LLM}{Large Language Model}
  \acro{LSTM}{Long Short-Term Memory}
  \acro{RMSE}{Root Mean Square Error}
  \acro{RNN}{Recurrent Neural Network}
  \acro{tanh}{hyperbolic tangent}
  \acro{VMM}{Vector Matrix Multiplication}
  \acro{VR}{Virtual Reality}
\end{acronym}

\bibliographystyle{IEEE}

\end{document}